\documentstyle[12pt,aasms4]{article}
\slugcomment{To appear in ApJ, October 10, 1997}

\lefthead{Renzini}
\righthead{Iron as a Tracer in Clusters}

\begin{document}

\title{Iron as a Tracer in Galaxy Clusters and Groups}

\author{Alvio Renzini\altaffilmark{1} 
\affil{European Southern Observatory, \\ Karl-Schwarzschild-Strasse 2,
D-85748, Garching bei M\"unchen, Germany, \\ arenzini@eso.org}}

% Notice that each of these authors has alternate affiliations, which
% are identified by the \altaffilmark after each name.  The actual alternate
% affiliation information is typeset in footnotes at the bottom of the
% first page, and the text itself is specified in \altaffiltext commands.
% There is a separate \altaffiltext for each alternate affiliation
% indicated above.

\altaffiltext{1}{On leave from: Dipartimento di Astronomia, Universit\`a di
Bologna, Italy.}

\catcode`\@=11
\def\gsim{\ifmmode{\mathrel{\mathpalette\@versim>}}
    \else{$\mathrel{\mathpalette\@versim>}$}\fi}
\def\lsim{\ifmmode{\mathrel{\mathpalette\@versim<}}
    \else{$\mathrel{\mathpalette\@versim<}$}\fi}
\def\@versim#1#2{\lower 2.9truept \vbox{\baselineskip 0pt \lineskip
    0.5truept \ialign{$\m@th#1\hfil##\hfil$\crcr#2\crcr\sim\crcr}}}
\catcode`\@=12
\def\ref{\par\noindent\hangindent=1truecm}
\def\yr-1{\hbox{${\rm yr}^{-1}$}}
\def\pd#1#2{\partial #1\over {\partial #2}}
\def\msun{\hbox{$M_\odot$}}
\def\mg2{\hbox{${\rm Mg}_2$}}
\def\mgas{\hbox{$M_{\rm gas}$}}
\def\ms{\hbox{$M_*$}}
\def\zfe{\hbox{$Z^{\rm Fe}$}}
\def\zfes{\hbox{$Z^{\rm Fe}_*$}}
\def\zfecm{\hbox{$Z^{\rm Fe}_{\rm ICM}$}}
\def\mout{\hbox{$M_{\rm out}$}}
\def\min{\hbox{$M_{\rm in}$}}
\def\als{\hbox{$\alpha_{\ast}$}}
\def\vtsn{\hbox{$\vartheta_{\rm SN}$}}
\def\lgm{\hbox{$L_{\rm grav}^{-}$}}
\def\lgp{\hbox{$L_{\rm grav}^{+}$}}
\def\lx{\hbox{$L_{\rm X}$}}
\def\feh{\hbox{${\rm [Fe/H]}$}}
\def\lsn{\hbox{$L_{\rm SN}$}}
\def\rsun{\hbox{$R_\odot$}}
\def\mast{M_{*}}
\def\mto{\hbox{$M_{\rm TO}$}}
\def\mf{\hbox{$M_{\rm f}$}}
\def\micm{\hbox{$M_{\rm ICM}$}}
\def\asn{\hbox{$\alpha_{\rm SN}$}}
\def\log{\hbox{${\rm Log}\, $}}
\def\rcs{r_{c*}}
\def\rch{r_{ch}}
\def\rt{r_{t}}
\def\re{r_{\rm e}}
\def\ie{I_{\rm e}}
\def\rsr{\rho _{*}(r)}
\def\rst{\rho _{*}}
\def\rhr{\rho _{h}(r)}
\def\ros{\rho _{0*}}
\def\roh{\rho _{0h}}
\def\mse{\widetilde m_{*}(\eta)}
\def\msd{\widetilde m_{*}(\delta)}
\def\mheb{\widetilde m_{h}(\eta,\beta)}
\def\mhdb{\widetilde m_{h}(\delta,\beta)}
\def\fiso{\widetilde \phi _{*}(0)}
\def\fise{\widetilde \phi _{*}(\eta)}
\def\fisd{\widetilde \phi _{*}(\delta)}
\def\fihob{\widetilde \phi _{h}(0,\beta)}
\def\fiheb{\widetilde \phi _{h}(\eta,\beta)}
\def\fihdb{\widetilde \phi _{h}(\eta,\beta)}
\def\msol{M_{\odot}}
\def\lsol{L_{\odot}}
\def\lb{\hbox{$L_{\rm B}$}}
\def\lt{\hbox{$L_{\rm T}$}}
\def\lgp{L^{+}_{\rm grav}}
\def\lgm{L^{-}_{\rm grav}}
\def\ho{\hbox{$H_\circ$}}
\def\h50{\hbox{$\ho /50$}}
\def\mstar{M_{\star}}
\def\egrav{{\widetilde {\cal E}}^{\pm}(\beta,\gamma,\delta)}
\def\egram{{\widetilde {\cal E}}^{-}(\beta,\gamma,\delta)}
\def\egramr{{\widetilde {\cal E}}^{-}(R,\gamma,\delta)}
\def\egravr{{\widetilde {\cal E}}^{\pm}(R,\gamma,\delta)}
\def\lgrav{L^{\pm}_{\rm grav}}
\def\parn{\par\noindent}
\def\pn{\par\noindent}
\def\arcsh{{\rm arcsh}}
\def\bmez{{\beta^2\over 2}}
\def\cbge{{\widetilde {\cal C}} _{\beta\gamma}(\eta)}
\def\cbgd{{\widetilde {\cal C}} _{\beta\gamma}(\delta)}
\def\egrav{{\widetilde {\cal E}}^{\pm}(\beta,\gamma,\delta)}
\def\egram{{\widetilde {\cal E}}^{-}(\beta,\gamma,\delta)}
\def\egrap{{\widetilde {\cal E}}^{+}(\beta,\gamma,\delta)}
\def\egramh{{\widetilde {\cal E}}^{-}_{h}(\beta,\delta)}
\def\egraph{{\widetilde {\cal E}}^{+}_{h}(\beta,\delta)}
\def\egrams{{\widetilde {\cal E}}^{-}_{*}(\delta)}
\def\egraps{{\widetilde {\cal E}}^{+}_{*}(\delta)}
\def\fie{\widetilde \phi (\eta)}
\def\fio{\widetilde \phi (0)}
\def\fiso{\widetilde \phi _{*}(0)}
\def\fise{\widetilde \phi _{*}(\eta)}
\def\fisd{\widetilde \phi _{*}(\delta)}
\def\fihob{\widetilde \phi _{h}(0,\beta)}
\def\fiheb{\widetilde \phi _{h}(\eta,\beta)}
\def\fihdb{\widetilde \phi _{h}(\delta,\beta)}
\def\intod{\int _{0}^{\delta}}
\def\lneb{\ln\left(1+{\eta^2\over\beta^2}\right)}
\def\lndb{\ln\left(1+{\delta^2\over\beta^2}\right)}
\def\lnxb{\ln\left(1+{\xi^2\over\beta^2}\right)}
\def\lgp{L^{+}_{\rm grav}}
\def\lgm{L^{-}_{\rm grav}}
\def\lgpm{L^{\pm}_{\rm grav}}
\def\mse{\widetilde m_{*}(\eta)}
\def\msx{\widetilde m_{*}(\xi)}
\def\msd{\widetilde m_{*}(\delta)}
\def\mheb{\widetilde m_{h}(\eta,\beta)}
\def\mhxb{\widetilde m_{h}(\xi,\beta)}
\def\mhdb{\widetilde m_{h}(\delta,\beta)}
\def\mbge{\widetilde M_{\beta \gamma}(\eta)}
\def\mbgt{\widetilde M_{\beta \gamma}(\tau)}
\def\mbgd{\widetilde M_{\beta \gamma}(\delta)}
\def\rcs{r_{c*}}
\def\rch{r_{ch}}
\def\rt{r_{t}}
\def\rsr{\rho _{*}(r)}
\def\rscsi{\widetilde \rho _{*}(\xi )}
\def\rse{\widetilde \rho _{*}(\eta )}
\def\rhcsi{\widetilde \rho _{h}(\xi,\beta)}
\def\rheb{\widetilde \rho _{h}(\eta,\beta)}
\def\ros{\rho _{0*}}
\def\roh{\rho _{0h}}
\def\yr{hbox{\rm yr}}
\def\pn{\par\noindent}
\def\me{\hbox{$M_{\rm e}$}}
\def\dvr{\hbox{$\Delta v_{\rm r}$}}
\def\mr{\hbox{$M_{\rm r}$}}
\def\mgas{\hbox{$M_{\rm gas}$}}
\def\mto{\hbox{$M_{\rm TO}$}}
\def\mst{\hbox{$M_{\star}$}}
\def\asn{\hbox{$\alpha_{\rm SN}$}}
\def\als{\hbox{$\alpha_{\ast}$}}
\def\ros{\hbox{$\rho_{\ast}$}}
\def\roh{\hbox{$\rho_{\rm h}$}}
\def\mc{\hbox{$M_{\rm c}$}}
\def\mo{\hbox{$M_{1}$}}
\def\mt{\hbox{$M_{2}$}}
\def\mur{\hbox{$M_{\rm 1R}$}}
\def\mdr{\hbox{$M_{\rm 2R}$}}
\def\tc{\hbox{$T_{\rm c}$}}
\def\mef{\hbox{$M_{\rm H\!eF}$}}
\def\yms{\hbox{$Y_{\rm MS}$}}
\def\ttr{\hbox{$t_{\rm tr}\;$}}
\def\tgwr{\hbox{$t_{\rm GWR}\;$}}
\def\3/2{\hbox{${3\over 2}$}}
\def\rpn{\hbox{$R_{\rm PN}\;$}}
\def\vexp{\hbox{$v_{\rm exp}\;$}}
\def\mer{\hbox{$M_{\rm e}^{\rm R}$}}
\def\men{\hbox{$M_{\rm e}^{\rm N}$}}
\def\lsun{\hbox{$L_\odot$}}
\def\lb{\hbox{$L_{\rm B}$}}
\def\lv{\hbox{$L_{\rm V}$}}
\def\lx{\hbox{$L_{\rm X}$}}
\def\lsn{\hbox{$L_{\rm SN}$}}
\def\lgravp{\hbox{$L_{\rm grav}^{+}$}}
\def\lgrav-{\hbox{$L_{\rm grav}^{-}$}}
\def\ldiscr{\hbox{$L_{\rm dscr}$}}
\def\lsigma{\hbox{$L_{\sigma}$}}
\def\msun{\hbox{$M_\odot$}}
\def\mfe{\hbox{$M_{\rm Fe}$}}
\def\mfecm{\hbox{$M_{\rm Fe}^{\rm ICM}$}}
\def\mfes{\hbox{$M_{\rm Fe}^*$}}
\def\rsun{\hbox{$R_\odot$}}
\def\ss{\hbox{$\sigma_{*}$}}
\def\sss{\hbox{$\sigma_{\star}^2$}}
\def\yr-1{\hbox{${\rm yr}^{-1}$}}
\def\mh{\hbox{$M_{\rm H}$}}
\def\mbol{\hbox{$M_{\rm bol}$}}
\def\dtp{\hbox{$\Delta t_{\rm peak}$}}
\def\logrhoc{\hbox{${\rm Log}\,\rho_{\rm c}$}}
\def\klu{(Dordrecht: Kluwer), p. }
\def\IMLR{Fe$M/L$}
% The abstract environment prints out the receipt and acceptance dates
% if they are relevant for the journal style.  For the aasms style, they
% will print out as horizontal rules for the editorial staff to type
% on, so long as the author does not include \received and \accepted
% commands.  This should not be done, since \received and \accepted dates
% are not known to the author.

\begin{abstract}
Available X-ray  data are collected and organized concerning the iron and gas
content of galaxy clusters and groups, together with the optical
luminosity, mass and iron abundance of cluster galaxies.
Moving from such a restricted number  of cluster parameters several
astrophysical inferences are drawn. These include the evidence for
rich clusters having evolved without much baryon exchange with their
surrondings, and having experienced very similar star formation
histories. Groups are much gas-poor compared to clusters, and  appear
instead to have shed a major fraction of their original cosmic share
of baryons, which indicates that galaxy clusters cannot have formed
by assembling groups similar to the present day ones. It is argued
that this favors low-$\Omega$ universes, in which the growth of rich
clusters is virtually complete at high redshifts. It is also argued
that elemental abundances in clusters are nearly solar, which is
consistent with a similar proportion of supernovae of Type Ia and Type
II having enriched both the solar neghborhood as well clusters as a whole.
Much of the iron in clusters appears to reside in the intracluster
medium rather than inside galaxies, the precise ratio being a function
of the Hubble constant.
It appears that the baryon to star conversion in clusters has been
nearly as efficient as currently adopted for the universe as a
whole. Yet the metallicity of the clusters is $\sim 5$ times higher
than the global metallicity adopted for the nearby universe. It is
concluded that the intergalactic medium should have a metallicity
$\sim 1/3$ solar if stellar nucleosynthesis has proceeded in 
stars within field galaxies with the same efficiency as in stars within 
clusters of galaxies.
\end{abstract}

% The different journals have different requirements for keywords.  The
% keywords.apj file, found on aas.org in the pubs/aastex-misc directory, 
% contains a list of keywords used with the ApJ and Letters.  These are 
% usually assigned by the editor, but authors may include them in their 
% manuscripts if they wish. 


% That's it for the front matter.  On to the main body of the paper.
% We'll only put in tutorial remarks at the beginning of each section
% so you can see entire sections together.
% In the first two sections, you should notice the use of the LaTeX \cite
% command to identify citations.  The citations are tied to the
% reference list via symbolic KEYs.  We have chosen the first three
% characters of the first author's name plus the last two numeral of the
% year of publication.  The corresponding reference has a \bibitem
% command in the reference list below.
%
% Please see the AASTeX manual for a more complete discussion on how to make
% \cite-\bibitem work for you.   

\section{Introduction}

The X-ray observations of  clusters of galaxies have revealed the presence
of large amounts of iron and other heavy elements in the intracluster
medium (ICM) (Mitchell, Ives, \& Culhane 1975; Serlemitsos et
al. 1976; Mushotzky et al. 1996), thus  providing  direct  evidence that gas contaminated by
nucleosynthesis processes has been lost  by galaxies in the course
of their evolution.
The access to the ICM elemental abundances  has offered 
the opportunity to investigate several important phenomena, such as the
baryon circulation on various scales (from galaxies to clusters), 
the integral past supernova (SN) activity along with the relative role of
the two major SN types, the efficiency of gas to galaxies, stars, and metals
conversion at a cluster scale, etc. (e.g., Vigroux 1977; Matteucci \&
Vettolani 1988; Arnaud et al. 1992; Renzini et al. 1993, hereafter
RCDP; Loewenstein \& Mushotzky 1996).
Some of these issues have been extensively discussed in RCDP
(see also Renzini 1994) on the basis of empirical evidence 
coming almost exclusively from {\it Einstein} X-ray
observations. In more recent years a great deal of relevant new data have
become available from the {\it ROSAT} and {\it ASCA} satellites,
and in this paper some of the issues discussed by  RCDP are revisited, and some
new inferences are drawn.

Clusters are the largest objects on which chemical enrichment can be
thoroughly studied,
and offer the additional advantage of being perhaps the best example in
nature for which the closed box approximation may hold true. Pei \&
Fall (1995) have recently modeled the chemical evolution of damped
Ly${\alpha}$ absorbers, thus predicting the evolution of the global
star formation rate with redshift (cosmic time). 
Global star formation rates all the way to $z\simeq 1$ and beyond have been
empirically constructed by Madau et al. (1996) using data from the 
Canada-France Redshift Survey (Lilly et al. 1996), and found it in
remarkable agreement with the predictions of Pei \& Fall.
Yet, absorbers represent a minor fraction of the baryonic matter,
unlikely to be well approximated by the closed box model, and Pei \&
Fall allow for both inflow and outflow in their chemical evolution
model. Clusters of galaxies are to some extent complementary to 
Ly${\alpha}$ absorbers, in that they contain a large amount of heavy
elements, partly in the hot ICM, partly locked into stars, none of which
participates in absorbing quasar light. With a bottom-up approach, in
this paper the iron (and
metal) content of clusters and groups is used as a tracer of baryon
circulation, past star formation, and supernova enrichment at the
cluster scale, and beyond.

The paper is structured as follows: Section 2 presents a recollection
of literature data concerning the iron and gas content of X-ray
clusters and groups, that will form the factual basis for the considerations
to be developed in the subsequent sections. In Section 3 several
inferences are derived from the uniformity of these cluster properties,
including the evidence for clusters having experienced little
baryon-exchange with their surrondings, having turned gas into
galaxies with nearly constant efficiency, and having experienced very
similar star formation histories. Groups instead present a completely
different scenario, which suggests that present day clusters cannot
have formed by coalescence of groups similar to the present day ones.
Section 4 deals with the elemental relative abundances at the cluster
scale,
while Section 5 deals with the global metallicity of clusters versus that
of the present day universe as a whole, and with the
global star formation history of the universe. The main conclusions of
this paper are finally listed in Section 6.
% In the first two sections, you should notice the use of the LaTeX \cite
% command to identify citations.  The citations are tied to the
% reference list via symbolic KEYs.  We have chosen the first three
% characters of the first author's name plus the last two numeral of the
% year of publication.  The corresponding reference has a \bibitem
% command in the reference list below.
%
% Please see the AASTeX manual for a more complete discussion on how to make
% \cite-\bibitem work for you.   

\section{The Iron Mass to Light Ratio of Clusters and Groups}

Ciotti et al. (1991) introduced the concept of ICM iron mass to
light ratio (\IMLR) as the ratio $\mfecm/\lb$ of the total iron mass in
the ICM over the total optical luminosity of galaxies in the cluster, and
found values in the range $(0.7-1.6)\times 10^{-2}\msun/\lsun$ for the nearby 
clusters Virgo, Coma, and Perseus. On a wider database, Arnaud et
al. (1992) found $\mfecm\propto\lb$, and RCDP finally adopted 
$\mfecm/\lb =0.01-0.02\msun/\lsun$ as typical for rich clusters. 
This value has a moderate dependence on the Hubble constant, being
proportional to $h^{-1/2}$ (Renzini 1994); for the data
presented in this Section $h=1/2$ has been adopted. The \IMLR \ relates
two pieces of {\it fossil}
information: the integral amount of iron ejected by galaxies in the
course of their evolution, and the present luminosity of the old stellar
remnant of the population that preasumably produced the observed iron,
early in its evolution.

The iron mass in the ICM is obtained as the product $\zfecm\micm$ of the iron
abundance times the mass of the ICM gas. The iron abundance is generally
estimated at the center of the cluster, and one assumes the
ICM to be chemically homogeneous. This is the case for rich,
high temperature clusters, while radial gradients in iron
abundance have been reported for cooler clusters from {\it ASCA} data 
(Ohashi et al. 1995). In
principle, it is possible to obtain $\mfecm$ integrating radially
$\zfecm$ in $\rho dV$. However,  {\it ASCA}'s PSF does not allow a
good reconstruction of the cluster ICM density profile.
Such integration could be accomplished by combining {\it ASCA}  data
with those of other X-ray telescopes with better PSF.

Figure 1 shows $\mfecm/\lb$ as a function of the cluster total
luminosity. For $\lb\gsim 4\times 10^{11}\lsun$ the \IMLR \ appears indeed
to be constant, with very small scatter around an average of
$\sim 0.02\,\msun/\lsun$. However, for $\lb\lsim 4\times 10^{11}\lsun$ 
-- i.e., in poor groups
rather than rich clusters -- the \IMLR \ exhibits much smaller values,
with no good correlation with the cluster optical luminosity. 
Figure 2 shows the same $\mfecm/\lb$ data now plotted as a function of
the temperature of the X-ray ICM. It is now apparent that the \IMLR \ is
virtually constant all the way to temperatures down to $\sim 1$ keV, 
and then drops precipitously by almost three orders of magnitude
as the temperature decreases below 1 keV.
The separation between groups and clusters is striking in this
diagram. It is worth mentioning that groups shown in this figures
represent a biased sample in that they are selected for being detected
in X rays. Typically, spiral rich groups (similar to the Local Group)
are not detected in X rays, which may be due to them having an even
lower ICM content, lower ICM mass to light ratio, and therefore lower
\IMLR .

The drop of the derived \IMLR \ in poor clusters and groups can be traced
to a drop in both factors entering in its definition, i.e., in 
the iron abundance {\it and} in the ICM mass to light ratio. Figure 3 shows
indeed ICM mass to light ratio $\micm/\lb$ as a function of the
cluster optical luminosity. While rich clusters have
fairly constant ICM mass to light ratio, much lower values with large
dispersion are exhibited by poor clusters and groups with 
$\lb\lsim 4\times 10^{11}\lsun$. The ICM mass is typically measured
within a radius that scales with the optical radius of the cluster,
and therefore it is smaller for groups than for clusters. 
Figure 4 shows $\micm/\lb$ as a
function of the ICM temperature. Again, fairly constant values
($\micm/\lb\simeq 30\msun/\lsun$) are exhibited by clusters hotter
than
$\sim 1$ keV, while the ICM mass to light ratio drops precipitously
below $\sim 1$ keV.

Figure 5 shows the iron abundance $\zfecm$ as a function of the cluster 
optical luminosity. Again, rich clusters with $\lb\gsim 4\times 10^{11}\lsun$
exhibit a fairly constant iron abundance, at the level of $\sim 0.3$
solar. Poorer clusters instead show a large dispersion, with very low
values being reached among poorest clusters. The iron abundance is
finally shown as a function of
ICM temperature in Figure 6. Clusters hotter than $\sim 2.5$ keV have 
similar abundances, with very small dispersion. Going to cooler clusters
$\zfecm$ appears first to increase up to near solar abundance for $kT$
slightly in excess of 1 keV, and then drops precipitously to almost
zero below 1 keV, with a fairly strong abundance-temperature
correlation (see also Arimoto et al. 1997, hereafter AMIOR). 
The presence of this correlation shows that just larger statistical
errors at low $kT$ cannot account for the apparent large spread of abundances.
As extensively discussed by AMIOR, very low iron abundances are also
derived from X-ray data for the ISM of ellipticals, which all have 
temperatures $\lsim 1$ keV. Also among ellipticals the iron abundance appears
to correlate with temperature, the lower the temperature the lower the
estimated  abundance (e.g., Davis \& White 1996).

It is apparent from all these figures that the \IMLR, the ICM mass to
light ratio, as well as the iron abundance all correlate much stronger
with cluster temperature rather than with optical luminosity. Part of the
problem may be due to observational errors, as cluster optical luminosities
are difficult to determine. However, there may be other effects at
work to make so strong the correlation with temperature. Some will be
explored in the next sections.

\section{The Baryon Circulation in Clusters and Groups}

\subsection{Rich Clusters}

The constancy of the \IMLR \ {\it and} of abundance among clusters
give support to the notion that galaxy clusters neither lost baryons to, nor
acquired baryons from the outside in the course of their evolution. Were
clusters loosing gas at some stage (e.g., as a result of the ICM heating by
galactic winds or AGNs) then iron would be lost as well, and sizable cluster to
cluster differences in the \IMLR \ should arise. Seemingly, were clusters
to accrete pristine gas ($Z^{\rm Fe}\simeq 0$) in substantial amount
then iron would be
diluted, and cluster to cluster differences in $\zfecm$ would arise.
Of course, one may argue that all clusters had nearly the
same amount of baryon exchange with their surrondings, which however seems
a rather contrived requirement.
This latter empirical result is in qualitative agreement with the
theoretical prediction according to which the baryon fraction of
clusters
cannot change appreciably in the course of their evolution (White et
al. 1993; Evrard 1997). 

The constancy of the \IMLR, ICM mass to light ratio, and abundance altogether
suggest that the conversion of baryons into galaxies, stars and
finally metals took place with nearly the same efficiency and at the same
epoch
in all clusters. Indeed, the constancy of the \IMLR \ says that the same
amount of iron was ejected by galaxies per unit present stellar
population. The most direct interpretation is that there are no
appreciable cluster to cluster differences in the global stellar
initial mass function (IMF) and star formation history, unless
variations in IMF are rather precisely compensated by variations in
the distribution of stellar ages (again, a rather contrived
requirement). 
For example, a
difference by $\Delta x = 0.5$ in the IMF slope would imply a factor $\sim 4$ 
difference in the \IMLR \ (cf. RCDP). Differences from cluster to cluster
in the average age of the stars producing the present optical
luminosity of the cluster would also affect the \IMLR. 
Assuming the cluster luminosity evolution to be dominated by the
passive aging of the elliptical galaxy population ($L\propto\simeq
t^{-(4-x)/3}$), a factor of two
difference in average stellar age implies a similar difference in
luminosity, hence in the \IMLR.

As well known, most of the cluster light $\lb$ is provided by
elliptical galaxies and galactic spheroids. Moreover, they represent
an even larger fraction of the stellar mass content of clusters since
these passively evolving systems have larger $M/\lb$ ratios than star forming
galaxies.
It is now well established by several independent lines of evidence
that the bulk of stars in cluster ellipticals
formed at high redshift, i.e., $z\gsim 3$ (for a review see e.g.,
Renzini 1995). 
This comes from the tightness of
the color$-\sigma$ relation of ellipticals in nearby clusters (Bower, Lucey,
\& Ellis 1992), the tightness of the  distributions of elliptical
galaxies about their fundamental plane at low and high redshift
(Renzini \& Ciotti 1993; van Dokkum \& Franx 1996; Pahre, Djorgovski,
\& de Carvalho 1997), the
tightness of the Mg$_2-\sigma$ relation, again at low as well as high
redshift (Bender, Ziegler, \& Bruzual 1996), the existence of red
(elliptical) galaxies at high redshift (Hamilton 1985;
Aragon-Salamanca et al. 1993; Dickinson 1995), and the tightness of
the color-magnitude relation of cluster ellipticals at high redshift 
(Ellis et al. 1996). The possibility has been advocated of fine tuning
between age and metallicity effects conspiring to
preserve the tightness of these relations at low-$z$ while allowing
for a large age spread (Worthey, Trager, \& Faber 1995). This appears to be
ruled out by the observation that the tightness is preserved even for
clusters at large lookback times, while it would obviously be
destroyed if a large age spread was present among ellipticals
in low-$z$ clusters (Kodama \& Arimoto, 1997).

It is most straigtforward to identify in the
spheroidal populations the producers of the bulk of the ICM iron in
clusters, since most of rich cluster stellar luminosity
comes from ellipticals, S0s, and the bulges of early-type spirals
(which are also dominated by very old stellar populations,
cf. Ortolani et al. 1995; Jablonka \& Alloin 1995). If so, then the
bulk of iron had to be produced by stars and ejected from galaxies
at very early times, say at $z\gsim 2-3$, even if a major fraction of
iron could have been produced by late SNIa exploders (RCDP). This -- 
coupled with
a non-evolving baryon fraction of clusters (White et al. 1993) -- implies
an ICM iron abundance independent of redshift, all the way to very
high redshifts. Some evidence in this direction is indeed emerging,
though for the moment limited to moderate redshifts (cf. Figure 6).

Although it appears most natural to attribute to ellipticals and
spheroids the prime role as iron (metal) producers in clusters, yet other
alternatives have been entertained and are worth considering. 
Giant elliptical galaxies and large spheroids have a rather deep potential
well, hence ejecting large amounts of metals requires a great deal of
energy. Dwarfs instead de-gas much more easily than giants, having much
shallower potential wells, and therefore dwarfs (e.g., dSph galaxies)
are likely to have ejected more iron per unit present light than
giants. This is
indeed a common feature of galactic wind models for the formation of 
ellipticals and spheroids (e.g., Larson 1974; Arimoto \& Yoshii 1987;
Matteucci 1992). Clearly, the problem is whether there were enough
dwarfs to make an appreciable fraction of the ICM iron. It has
actually been argued that most ICM itself might have been ejected from
dwarfs, given that the dwarfest among dwarf spheroidals have a very low 
baryonic
fraction, hence had to eject most of their baryons if they were to start with
the cosmic share (Trentham 1994). However, both the faint end of the galaxy 
luminosity function, and the relative amounts of baryons and metals
ejected by dwarfs are sufficiently uncertain to make very hard either
to prove or disprove this scenario. 
Most baryons are now in the ICM rather than within
galaxies (i.e., galaxies contain $\sim 20\%$ of the baryons, e.g.,
David et al. 1990; White et al. 1993). Hence, the question is whether 
 the baryons were separated from
their (small scale) dark matter complement by local starbursts, or by
mutual stripping in the process of hierarchical clustering (Nath \& Chiba
1995), when small scale density peaks dissolved inside the collapsing cluster.
 In the former case a population of low metallicity stars would be present,
either  still clustered in e.g., dSph's, or dispersed through the whole cluster
potential well. 
However, a dispersed population can also arise from tidal stripping
of galaxies in the course of the cluster evolution (cluster {\it
harrassment}, cf. Moore et al. 1996), and it
will be very hard to  distinguish observationally between the two options.
Upper limits to a diffused stellar population in clusters have been 
reported to be at the level of $\sim 20\%$
of the total cluster light (Melnick, White, \& Hoessel 1977). A
promising way to better determine this fraction may come from
detecting and counting isolated planetary nebulae that are now being
found in clusters (Theuns \& Warren 1997; Arnaboldi et al. 1996).
Theuns \& Warren argue that perhaps as much as 40\% of the cluster 
stellar light comes from a dispersed population, but also point out
that this estimate is
based on an uncertain frequency of planetary nebulae per unit light of
the parent population.
This is known to vary by a very large amount among spheroidals, with
most
metal rich ones having a several times smaller PN productivity compared to 
metal poor ones (Hui et al. 1993; Ferguson \& Davidsen 1993).

Another possibility is offered by 
an early generation of massive stars in a now extinct (i.e.,
$\lb\simeq 0$) population produced in a {\it bimodal} star formation
scenario (Arnaud et al. 1992; Elbaz, Arnaud, \& Vangioni-Flam 1995).  
One postulates that only massive stars (e.g., $M\gsim 2\,\msun$) would
form in starbursts, while the low mass stars that shine today in
galaxy clusters would come only later in a {\it quiescent} star
formation mode. For this model to work fine tuning is required to
explain the constancy of the \IMLR, with the ratio of the star yields
of the two star formation modes being the same in all clusters.
Moreover, there is little evidence for bimodal star formation in
nature. For example, a typical globular cluster formed $\sim
10^6\msun$ of stars  in probably less than $10^6$ yr, with a star
formation rate of $\sim 1\,\msun\yr-1$ per cubic parsec (!), a strong
starburst indeed, and yet they were able to produce the wealth of low
mass stars without which we would not see them today. Moreover,
abundant low mass stars have been detected in Galactic regions with
active high mass star formation (Zinnecker, McCaughrean, \& Wilking
1993), and  therefore the occurrence of bimodal star formation  appears rather
implausible.
All in all, it appears natural to adopt the view that the metals we
see in clusters were produced by the same stellar population which low
mass component allows us to see cluster galaxies today.

\subsection{Galaxy Groups}

An inspection to Figures 1-6 immediately reveals that what holds for
clusters, either hotter than $\sim 2$ keV or brighter than $\sim
4\times 10^{11}\lsun$, apparently does not hold for cooler/fainter
clusters and groups. I first assume $\zfecm$ and $\micm/\lb$ ratios
for these latter objects at face value, and follow their astrophysical
implications. Some of such implications were already mentioned in
RCDP, even if at that time data were available for only one object,
namely the NGC 2300 group (Mulchaey et al. 1992). The very low values
of the \IMLR \ shown in Figure 2 imply that a great deal of iron
should have been lost by the groups, hence a major fraction of their
baryons along with it. This appears to be in agreement with the ICM
mass to light ratio being indeed much lower than in clusters (Figure
4). However, this cannot be the whole story, because besides a lower
\IMLR \ several groups appear to have a much lower iron abundances
compared to clusters (Figure 6). It seems as if, after having shed a
major fraction of their original baryon content, such groups had to
re-accrete a modest amount of baryons from their surrondings, where
pristine gas largely diluted the metal contaminated ejecta of the
groups themselves (RCDP). This alternate flow of baryons, out and in
groups, may arise as a result of a declining galactic wind activity
inside groups, and at least at early times such activity could well
have been strong enough to drive gas out of groups (Renzini
1994). However, the whole scenario appears rather contrived, as it may
be difficult for groups to reaccrete gas once its adiabat has been
raised by energy input from the galactic winds (Kaiser 1991).

There is however another interpretation still open. It is apparent
from
Figures 2, 4 and 6 that for $kT\gsim 2$ keV all plotted cluster properties
are nearly constant and exhibit very small scatter. Cooler clusters
and groups,
especially for $kT\lsim 1$ keV, show instead a wild range of
abundances, gas mass to light ratios, and therefore \IMLR \ values. It is worth
emphasizing that the diagnostic tools to get the iron abundance are
radically different in the hot clusters compared to the cooler
ones. Above $kT\simeq 2$ keV iron abundances  are derived primarily from the
iron-K complex
at $\sim 7$ keV, where lines arise from transitions down to the K
shell of He-like and H-like iron ions. The atomic physics used in this
plasma emission models is therefore rather simple. At cooler
temperatures instead, iron abundances are derived from the iron-L
complex at $\sim 1 $ keV, where lines arise from transitions down to
the
L shell of progressively more and more complex iron ions as
temperature becomes lower and lower. The more complex the ions, the
less secure the involved atomic physics calculations,
especially the collisional excitation probabilities, and it appears
legitimate to entertain the suspicion that much of the observed trends
below $\sim 2$ keV may just be an artifact of some systematic error
in the iron-L diagnostics (cf. AMIOR for an extensive discussion). 
In the case of the Virgo cluster ($kT\simeq 2.9$ keV) it has been shown
that the iron
abundance derived from iron-L lines is in very good agreement with
that derived from iron-K lines (AMIOR).  Hwang et al. (1997) come to
the same conclusion analysing a sample of clusters with $kT$ in the 2-4
keV temperature range. Unfortunately, by no means the reliability of the iron-L
diagnostics at $kT\simeq 3$ keV can prove its reliability at
lower temperatures, since then other, much more complex iron ions are
involved (AMIOR). In conclusion, the drop of iron abundance below
$\sim 1$ keV may not be real at all, and therefore it is premature to
conclude that groups had to re-accrete baryons at late epochs. 
It appears that the best way to assess the reliability of iron-L
diagnostics at low temperatures will be offered by collecting ASCA
observations of objects for which the iron abundance is independently
known, such as stars, supernova remnants, starburst galaxies etc. 
(cf. AMIOR for a preliminary discussion).

While the iron abundances may be in error at low $kT$ values, it is
worth emphasizing that the low $\micm/\lb$ ratios shown in Figure 4
should be only marginally affected by possible systematic errors in
the iron-L diagnostics, i.e., while abundances and \IMLR \ values may be
severely underestimated below $\sim 1$ keV, the low values of the 
ICM mass to light ratios in groups are more robust, and one can safely
conclude that groups are genuinely gas poor, likely
to have lost a major part of their original share of baryons.
Alternatively, in groups baryon conversion to stars  would have been much more
efficient than in clusters, proceeding almost to the last ``drop'' of
them,
which seems a less plausible interpretation.

\subsection{Clusters vs. Groups and the Formation Epoch of Clusters}

The systematic difference in the $\micm/\lb$ ratio of groups relative
to rich clusters can have far reaching consequences. Indeed, for merging
groups the $\micm/\lb$ ratio should stay the same or decrease
slightly (Evrard 1997), therefore failing to meet the high $\micm/\lb$ values
typical of clusters, which are up to $\sim 30$ times higher than those
in groups.
This conclusion would be further reinforced were the iron-L diagnostics
correct, since merged groups would also have much lower \IMLR \ 
compared to clusters.

The inference is that rich clusters cannot have formed by merging
groups similar to those in the nearby universe. If clusters formed by 
hierarchical merging, merging should have taken place when groups were
still gas rich, likely at high redshift.  This favors low-$\Omega$
universes, where little merging and cluster growth takes place at late
epochs (low $z$), while most of merging activity ``switches off'' at
$1+z\simeq\Omega_\circ^{-1}\simeq 5$ in an open universe, or at
$1+z\simeq\Omega_\circ^{-1/3}\simeq 1.5$ in a $\Lambda$CDM universe
(White 1997). Late accretion of groups onto rich clusters may still
take place, but should occur at such a low rate to leave substantially
unaffected the ICM mass to light ratio of the clusters.
Independent evidence for very little or no cluster evolution all the
way out to $z\simeq 0.6$ is now emerging both from optical (Carlberg et
al. 1996) as well as X-ray selected cluster samples (Rosati et al. 1997).

This conclusion is further reinforced by some of the considerations
in Section 3.1. Indeed, with most of stars in clusters having formed
at high redshift ($z\gsim 3$), and with the concomitant production and
ejection of iron, much of the heating of the ICM  also took place at such
early epoch (e.g., Renzini 1994). If clusters were not already in
place such heating would have drasticaly affected the baryon content
of clusters, as it appears indeed to have done for groups. While small
cluster to cluster variations in the baryonic fraction might have been
detected (Loewenstein \& Mushotzky 1996b), such variations are tiny
compared to the range covered by the $\micm/\lb$ ratio of groups in
Figures 3 and 4. Rich clusters appear to be ``older'' than the bulk of
the old stellar population by which they are dominated.

However, one may also envisage a scenario in which 
groups destined to assemble forming a rich cluster  expelled gas at an
early stage, the gas remained confined as an intragroup medium, and
finally was collected inside the cluster along with the groups. 
%This
%might be a viable alternative provided the low iron abundances
%indicated by the iron-L diagnostics are incorrect. Otherwise, the
%intragroup medium would be iron poor compared to the observed ICM.

\section{The Elemental Ratios in the Intracluster Medium}

It is now well established that in the metal poor stars of the
Galactic halo the $\alpha$-elements are enhanced with respect to iron
relative to the solar proportions  (e.g.,
Wheeler, Sneden, \& Truran 1989; Bessell, Sutherland, \& Ruan 1991),
and a similar enhancement is observed in the metal rich stars of the
Galactic bulge (McWilliam \& Rich 1994). The current interpretation of
this $\alpha$-element overabundance in  the whole Galactic spheroid
appeals to the prompt release of $\alpha$-elements by short living
massive stars producing Type
II SNs, coupled to the somewhat delayed release of a major fraction of
the whole iron by Type Ia SNs (e.g., Greggio \& Renzini 1983;
Matteucci \& Greggio 1986; Ruiz-Lapuente, Burkert, \& Canal 1996).
Various degrees of $\alpha$-element enhancement can thus be obtained,
depending
on the adopted time scale of the iron release by SNIa's relative to
the time scale of star formation. This scenario is referred to as the
``standard chemical model'' for the Galactic chemical evolution (RCDP).
Moreover, population synthesis methods indicate that an
$\alpha$-element enhancement  may also be present in ellipticals,
reaching up to 0.2-0.3 dex in the most massive ones (e.g., Worthey,
Faber, \& Gonzales 1992; Davies, Sadler,
\& Peletier 1993). By analogy with the Galactic spheroid, the favored
interpretation has been in
terms of the relative roles of the two SN types during the fast
formation process of ellipticals.

A fast completion of star formation (on time scales of, say $\lsim$ few
$10^8$ yr) implies that a fraction of the iron released by SNIa's
should flow directly out of galaxies during an early galactic wind
phase, without  ever being incorporated into stars. If so, a {\it
chemical asymmetry} should be established between galaxies and the
ICM, with galaxies being slightly overabundant in $\alpha$-elements 
relative to iron, and the ICM being slightly overabundant in iron
relative to the $\alpha$-elements (RCDP). The size of this asymmetry
is hard to predict theoretically, as it depends on the relative
time scales of star formation on the one hand, and of the SNIa iron
release on the other, with both time scales being  poorly known. 
The effect is not expected to be large. If all
the iron from SNIa were ejected from galaxies the $\alpha$-deficiency
in the ICM would be [$\alpha$/Fe]$\sim -0.4$ (RCDP). The actual
deficiency may  be much smaller than this, since part of the
SNIa iron is likely to be incorporated into stars, while SNII products
are also ejected from galaxies.

Early attempts at measuring the abundance of $\alpha$-elements in the
ICM indicated a possible overabundance of oxygen relative to iron
(Canizares et
al. 1982) and a near solar Si/Fe ratio (Mushotzky et al. 1981),
but both with large statistical and systematic uncertainty.
{\it ASCA} has now provided much better data, and Mushotzky (1994) has
initially reported a fairly high $\alpha$-element enhancement, with
$<\![\alpha$/Fe]$>\simeq +0.4$ (from his Table 3). More recently
Mushotzky et al. (1996) have revised down this estimate, and report
detailed elemental abundances for several $kT=$3--4
keV clusters, all showing a moderate $\alpha$-element enhancement.
Uncertainties in the abundances are reported to be of about a factor
of two, at the 90\% confidence level.
Taking a global average for O, Ne, Mg, Si, and Fe, one obtains a modest
$<\![\alpha$/Fe]$>\simeq +0.2$. This is still in conflit with the
predicted asymmetry, and argues for not only the galaxies, but the ICM
as well being dominated by SNII products (Loewenstein \& Mushotzky
1996a). If so, this would
demonstrate the impossibility to extend to galaxy clusters the
standard chemical model that apparently holds for the Galactic
nucleosynthesis (AMIOR).

However, Ishimaru \& Arimoto (1997) have recently pointed out that the
small $\alpha$-element enhancement in the ICM comes from Mushotzky et
al. (1996)
having assumed reference solar abundances from ``photospheric'' model
atmosphere analysis. The result is different if one uses ``meteoritic'' 
abundances instead, because the meteoritic iron abundance is $\sim 0.16$
dex  lower than the photospheric value (Anders \& Grevesse
1989). After noting that the meteoritic iron is now more generally adopted, 
Ishimaru \& Arimoto conclude that the $\alpha$-element enhancement in
the ICM virtually disappears, going down from $\sim $0.2 to only $\sim 0.04$
(consistent with no enhancement or even small depletion, given the reported
errors).
 
At this stage it is safe to conclude that there is no strong evidence for an
enhanced $\alpha$-element proportion in the ICM, and that the
galaxy-ICM chemical asymmetry -- if it exists -- must be small and
probably hard to detect given the current errors in abundance
determinations. Abundance ratios in the cluster as a whole (including
stars and ICM together) are therefore very close to solar, as expected
in the frame of the standard chemical model (RCDP). Therefore, the
applicability of the standard chemical model to clusters is not
invalidated, and the 
possibility of a major contribution from SNIa's to the iron enrichment
of clusters remains viable.

\section{The Metallicity of the Present Day Universe and the Past
History of Star Formation}

Clusters of galaxies are the largest entities for which we can
measure the metallicity. It is important to notice that the mass of
iron in the ICM is comparable to
that locked into galaxies, having assumed the average abundance of stars in
galaxies to be solar (RCDP). While the abundances in either the ICM or
galaxies are independent of the distance scale, the cluster average
abundance does actually depend on the assumed Hubble constant, because
so do both the mass in galaxies and $\micm$, and each of them in a
different way. I assume as prototypical the Coma cluster values
adopted by White et al. (1993): $\micm\simeq 5.5\times
10^{13}h^{-5/2}\msun$ and $M_*\simeq 10^{13}h^{-1}\msun$.
 and derive for the cluster iron abundance:
$$Z_{\rm CL}^{\rm Fe}={\zfecm\micm + \zfes M_* \over \micm + M_*}=
      {5.5\zfecm h^{-5/2} + \zfes h^{-1}\over 5.5h^{-5/2} +
      h^{-1}},\eqno(1)$$
where $\zfes$ is the average abundance of stars in galaxies and $ M_*$
is the mass in stars. With $\zfecm=0.3$ solar and $\zfes=1$ solar,
equation (1) gives a global cluster abundance of 0.34,
0.37, and 0.41 times solar, respectively for $h=0.5$, 0.75, and 1. 
Under the same assumptions, the ratio of the iron mass in the ICM to
      the iron mass locked into stars is:
$${\zfecm\micm\over\zfes M_*}\simeq 1.65 h^{-3/2},\eqno(2)$$
or 4.6, 2.5, and 1.65, respectively for $h=0.5$, 0.75, and 1. Note
that with the adopted values for the quantities in equation (2) most
of the iron is in the ICM, rather than now locked into stars,
especially for low values of $\ho$. These estimates could be somewhat decreased
if clusters contain a sizable population of stars not bound to
censed individual galaxies (cf. Section 3.1). The iron content of
galaxies may also have been underestimated, because so does a
luminosity-weighted abundance compared to the mass-weighted abundance
(e.g., Greggio 1997). In addition, the galaxy (baryonic) $M_*/L$ ratio
might be 
higher than estimated by White et al. (1993), i.e.,
$<\!M_*/\lb\!>=6.4h$. All these effects
together may contribute to reduce the ICM to galaxies iron ratio below
the somewhat embarrasingly large values given above.

With the adopted masses and iron abundances for the two baryonic
components  one can also evaluate the total
cluster \IMLR. One needs to specify the average $M_*/\lb$ ratio, and
adopting for consistency the value given by White et al. (1993),
one gets:
$${\mfecm +\mfes\over\lb}\simeq 1.3\times 10^{-2}(1.65\, h^{-1/2}+h)
\quad (\msun/\lsun),\eqno(3)$$
or \IMLR=0.037 or 0.034 $\msun/\lsun$, respectively for $h=0.5$ and
1. The total
\IMLR \ is therefore fairly insensitive to the adopted distance scale,
and is close to the value previously estimated (i.e.,
0.03$\msun/\lsun$, RCDP; Renzini 1994). Simple calculations (cf. RCDP)
show that to reproduce this value one needs either
a fairly flat IMF ($x\simeq 0.9$) if all iron is attributed to SNII's,
or a major contribution from SNIa's, if one adopts a Salpeter IMF
($x=1.35$). The former option
dictates a substantial $\alpha$-element enhancement, similar to the
values observed in the Galactic halo ([$\alpha$/Fe]$\simeq +0.5$).
The latter option instead predicts near solar proportions for the
cluster as a whole. On the basis of the discussion in Section 4 one
concludes in favor of the second option.
Moreover, the total metal mass to light ratio of clusters is 
$\sim 0.3\,\msun/\lsun$, given that in the solar proportion iron
accounts for about 10\% of all metals. There is no doubt that the bulk
of such metals other than iron -- hence of the metals as whole -- was
produced by SNIIs. Now, the number
of SNIIs exploded at early times in a stellar population that has
faded to luminosity $\lb$ when aged to $\sim 15$ Gyr is $\sim
0.1\times\lb /\lsun$ (RCDP). Therefore, to produce the observed metal mass to
light ratio of clusters each SNII must have contributed on average 
$\sim 3\msun$ of heavy elements, which is in agreement with current
nucleosynthesis calculations (e.g., Woosley \& Weaver 1995).

A critical issue is to what extent the cluster global metallicity, and
the ICM to galaxies iron share are representative of the low$-z$
universe as a whole. For example, Madau et al. (1996) adopt $\ho=50$, a stellar
mass density parameter $\Omega_*=0.0036$, a baryon mass density
parameter $\Omega_{\rm b}=0.05$, an average solar metallicity for the
stars, and a negligible metal content for the intergalactic medium (IGM),
that comprises the vast majority of the baryons. With these
assumptions the metallicity of the present day universe is $\sim
1\times 0.0036/0.05= 0.07$ solar, or $\sim 5$ times lower than the measured
value in clusters of galaxies. In the same frame,  the fraction of baryons in
galaxies (stars) is also $\sim 7\%$ globally, which compares to  $\sim
1/(1+5.5h^{-3/2})$ in clusters, or $\sim 6\%$ and $\sim 10\%$,
respectively for $h=0.5$ and 0.75.
Therefore, it appears that the efficiency of baryon conversion into
galaxies and stars ($\Omega_*/\Omega_{\rm b}$) adopted by Madau et
al. (1996) is nearly the same as that observed in clusters, which
supports the notion of clusters being representative of the low-$z$
universe. The metallicity of the clusters is however $\sim 5$ times higher
than the metallicity of the low-$z$ universe as conservatively 
adopted by Madau et al.. 
The difference  comes from having attributed
to the IGM a very low metallicity, hence assuming field galaxies losing only
a negligible amount of metals.  This implies a factor $\sim 5$ lower efficiency
in metal production per unit mass turned into stars. I will discuss
the two aspects in turn.

Is there any reason why most of the  produced metals should be ejected by 
cluster galaxies, and instead fully retained by field galaxies, while
reaching the same average stellar metallicity? The only hint for a
systematic cluster/field difference may be offered by ram pressure
stripping, that to some extent should be active in clusters but not in
the field.
However, in clusters iron was most likely
ejected from galaxies as a result of supernova heating, rather than
stripped by ram pressure (cf. RCDP). In particular, ram pressure
stripping cannot have played a major role because  as shown by Figure
2 the \IMLR \ is the same in clusters with moderate velocity dispersion (or 
equivalently, $kT\simeq 2$ keV) as in clusters with high velocity
dispersion (or $kT\simeq 10$ keV). Were ram pressure important one
would have expected the \IMLR \ to increase with ICM temperature.
There appears also to be no strong argument for field galaxies
having shed much less iron than cluster galaxies. The small starburst
galaxies making the excess blue galaxy counts (Lilly et al. 1995) are
indeed likely to have ejected a sizable amount of metals before
``fading to oblivion''.

Seemingly, there are no strong arguments supporting the factor of $\sim 5$
difference in metal production efficiency, unless one is willing to
postulate a flatter stellar IMF in cluster relative to field
galaxies. The difference might be
somewhat smaller if one allows for the possibility that there are more
stars in clusters than  adopted here (cf. Section 3.1). However, if such stars
escaped detection within well studied clusters, they may have
gone unnoticed in the field as well. Therefore, it would appear quite
{\it ad hoc} to appeal to a much higher metal productivity of stellar
populations in galaxy clusters compared to their field counterpart,
and it appears reasonable to conclude that the global
metallicity of the present day universe may well be nearly the same as
that observed in galaxy clusters, i.e., $\sim 0.3-0.4$ times solar.
If so, there should be a comparable share of metals in the field IGM,
as there is in the cluster ICM.

While it appears that the global, time-averaged rate
of metal production in the universe may have been
underestimated by perhaps as much as a factor $\sim 5$, this does not
necessarily imply that the   global, time-averaged rate of star
formation was also underestimated by the same factor.
More simply, the metal productivity of galaxies per unit of their
present mass may have been underestimated by Madau et al., if the
cluster galaxies productivity is to be taken as representative of the
universe as a whole. One cannot  exclude that actual average rate of
star formation may have been underestimated, which would be the case if
a population exists of unaccounted stars in clusters and in the field.

\section{Conclusions}

In this paper several relations  coming from combining galaxy
cluster X-Ray and optical data have been presented and discussed, 
and astrophysical inferences have been drawn from them. The main results can
be summarized as follows.\pn
1. The ICM iron mass to optical light ratio of clusters with $kT\gsim 2$ keV
appears
to be constant, with a small dispersion which is fully consistent with
observational errors. Among these clusters also the iron abundance and
the ICM mass to light ratio appear to be constant with small dispersion.
\pn
2. Among poor clusters and groups, instead, the \IMLR, the ICM mass to
light ratio, and the iron abundance all drop by orders of
magnitude
compared to the corresponding values typical of the rich clusters.
This is especially evident for ICM temperatures  below $\sim 1$ keV.
\pn
3. The constancy of the \IMLR \ among clusters indicates that clusters
did not lose an appreciable fraction of their original share of the
cosmic
baryons. It also argues for the  stellar initial mass function
being
nearly the same in all clusters, as well as for the average stellar
ages being also the same from cluster to cluster, within less than 
a factor $\sim 2$.
\pn
4. As a major fraction of the cluster optical luminosity comes from ellipticals
and bulges, the stellar population responsible for the production of
the iron and the other heavy elements now in the ICM is most naturally
identified with the high mass tail of the same population now
surviving in ellipticals and bulges. Since such spheroid populations
formed at high redshift ($z\gsim 3$), much of the iron should have
been ejected from galaxies at a very early stage. This is in agreement with
the observed iron abundance in moderate redshift clusters being the
same as in local clusters.
\pn
5. The possibility remains that a non-negligible fraction of the
cluster stellar population resides out of censed galaxies. If so, a
corresponding fraction of the ICM metals could have been manufacted by
this dispersed stellar population. In principle, part of the metals
could have been produced by a now extinct stellar population,
generated in a bimodal star formation mode. It is argued that this
hypothesis is rather implausible.
\pn
6. The drop of the ICM mass to light ratio in groups indicates that
-- contrary to clusters --
they have experienced a major loss of gas (baryons) in the course of
their evolution. The very low metallicity indicated by current X-ray
diagnostics for most
of these groups, if real, suggests that baryons should have been
re-accreted after an initial de-gassing, which seems rather contrived.
However, it is possible that current iron-L diagnostics for $kT\lsim
1$ keV is affected by systematic errors.
\pn
7.
The large differences between the cluster and the group ICM mass to
light ratios suggest  that clusters did not form by agglomerating
groups similar to the present day ones. If such agglomeration took place, it
must have occurred while the groups were still gas rich, i.e., before 
major star formation took place thus propelling part of the
baryons out of them. This argues for the assembly of rich clusters
having been completed at high redshift, which favors low-$\Omega$
universes.\pn
8. There is not much evidence for a predicted chemical asymmetry in
the
$\alpha$-element to iron ratios
between the ICM and the stellar component of clusters. However, data
are consistent with no $\alpha$-element enhancement in the ICM,
and therefore with a combined SNIa and SNII enrichment to give near
solar elemental proportions on the whole cluster scale, as predicted
by the standard model for the Galactic chemical evolution.
In particular, there appears to be no need to invoke a special IMF or
a suppression of SNIa's in clusters, compared to our own Galaxy.
\pn
9. It is emphasized that the overall cluster metallicity is about 5
times higher than the currently adopted average metallicity of the
present-day
universe, in spite of a similar fraction of the baryons having been
converted into galaxies and stars in the clusters as well as in the 
general field. Possible origins of the difference (or discrepancy) are
discussed, including an underestimate of the efficiency of metal
production per unit mass of baryons converted into stars.
\pn
10. The  amount of iron (metals) in an
undetected, probably hot intergalactic medium comprising most of the
bayons is predicted to be comparable to -- or
even larger than -- the amount of iron now locked into stars in the
present day universe. The metal abundance of this IGM is predicted to
be $\sim 1/3$ solar, if the metal productivity of stellar populations
is the same in the field  as it is in clusters. 
\bigskip\pn
I am grateful to Mauro Giavalisco and Piero Madau for their 
support and for useful
discussions about the metallicity of the present day universe and the
integrated past star formation activity, and to Ralph Bender and an
anonymous referee for
constructive comments. I wish to thank Richard Ellis
and Trevor Ponman for stimulating discussions, and for their invitation to
participate at the Royal Astronomical
Society meeting of October 11, 1996, and the Pontificial Academy of
Sciences for its invitation to attend the Vatican Conference on ``The
Emergence of Structure in the Universe at the Level of Galaxies'', 
November 25-29, 1996, as at these meetings  I had the opportunity to
report and discuss in a friendly environment the results now presented
in this paper.

\clearpage

\clearpage

\figcaption[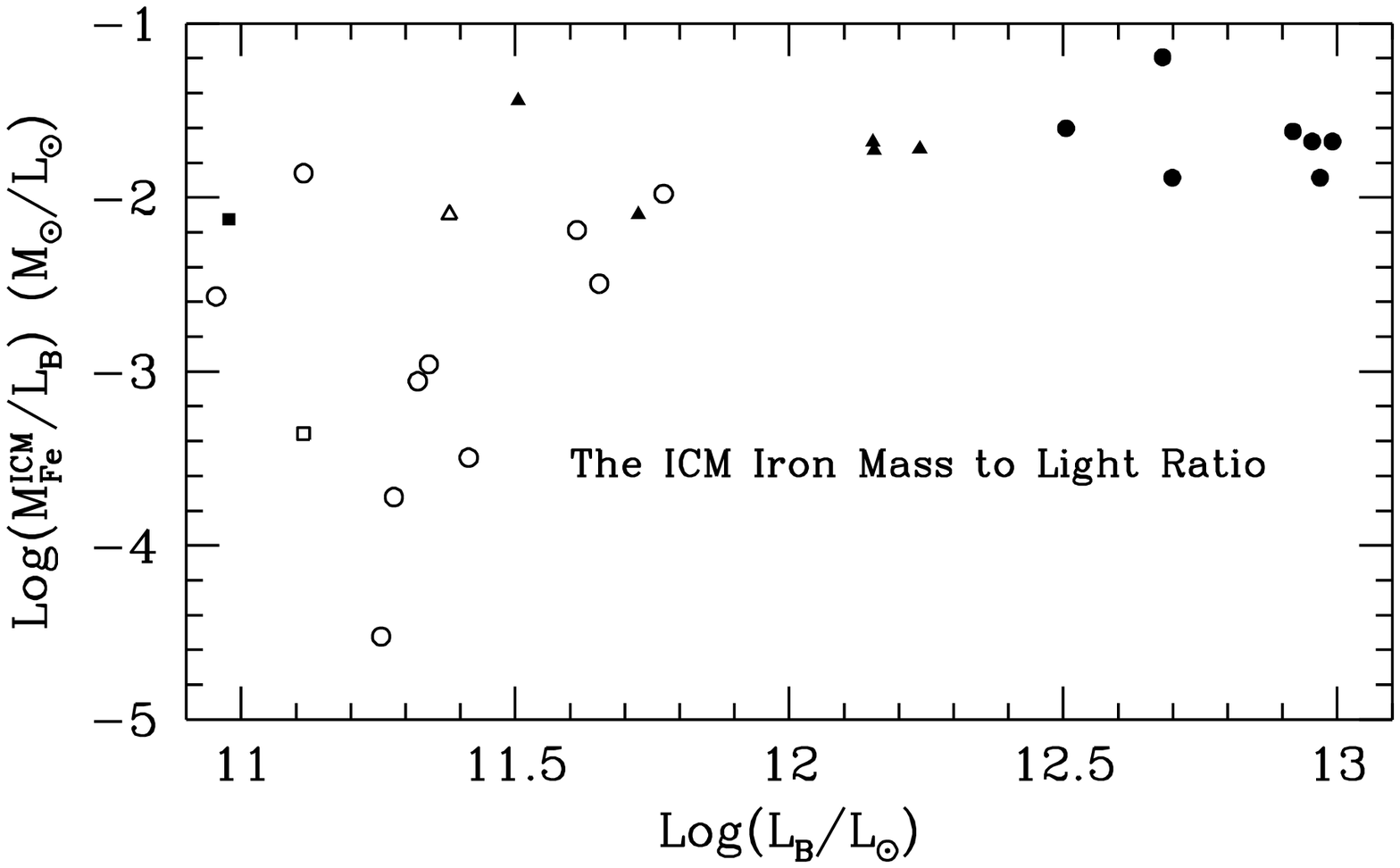]{The iron mass to light ratio  of  the  ICM
of clusters
and groups as a function of the total optical luminosity $\lb$ of the
cluster galaxies. 
Data are taken from the following sources: filled circles: Arnaud
et al. (1992); filled triangles: Tsuru (1993); open triangle: David   et al.
(1994a); open square: Mulchaey et al. (1993); filled square:
Ponman et al. (1994); open circles: Mulchaey et al. (1996). 
 \label{fig1}}

\figcaption[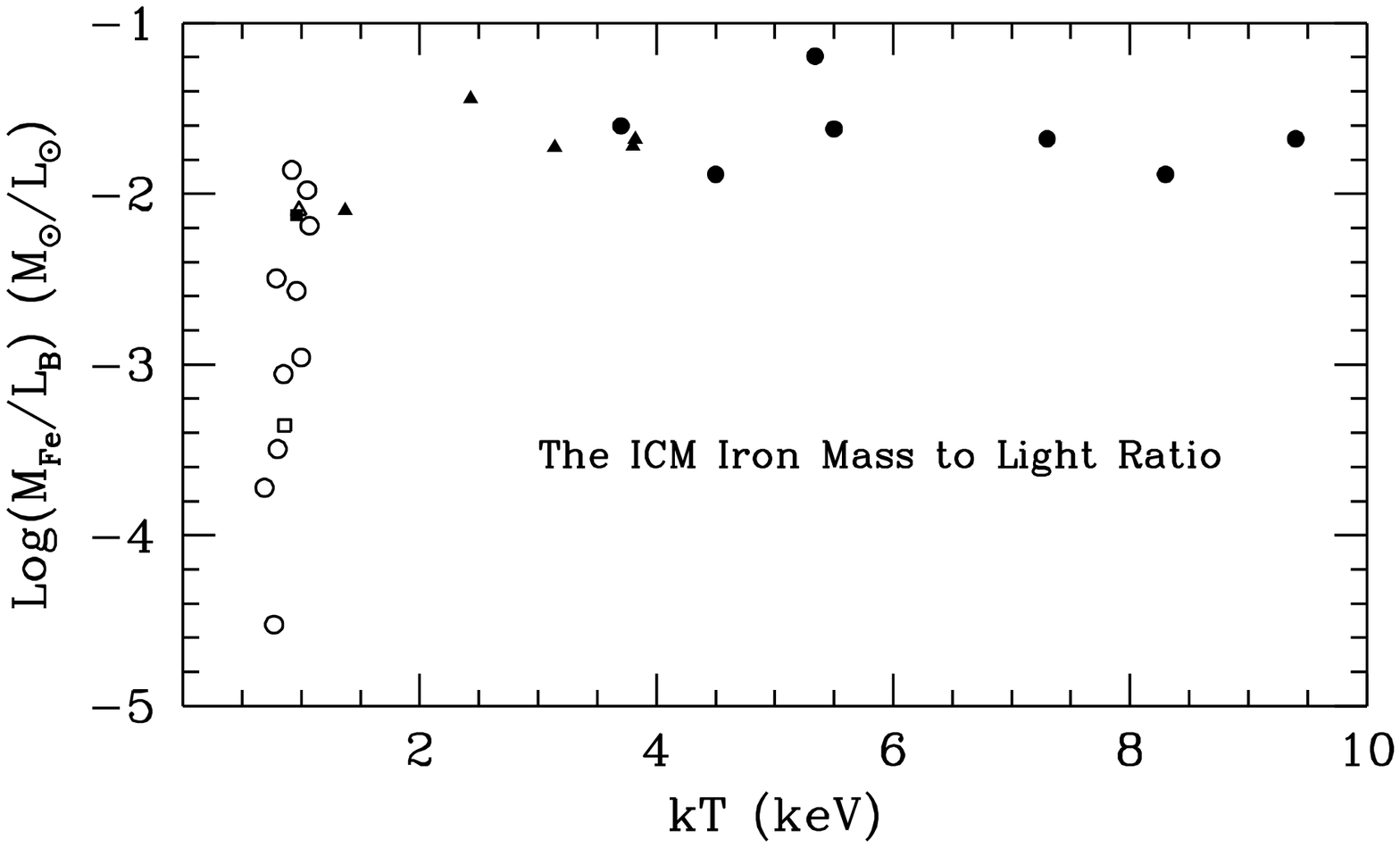]{The same as Figure 1 but as a function of
the ICM temperature. \label{fig2}}

\figcaption[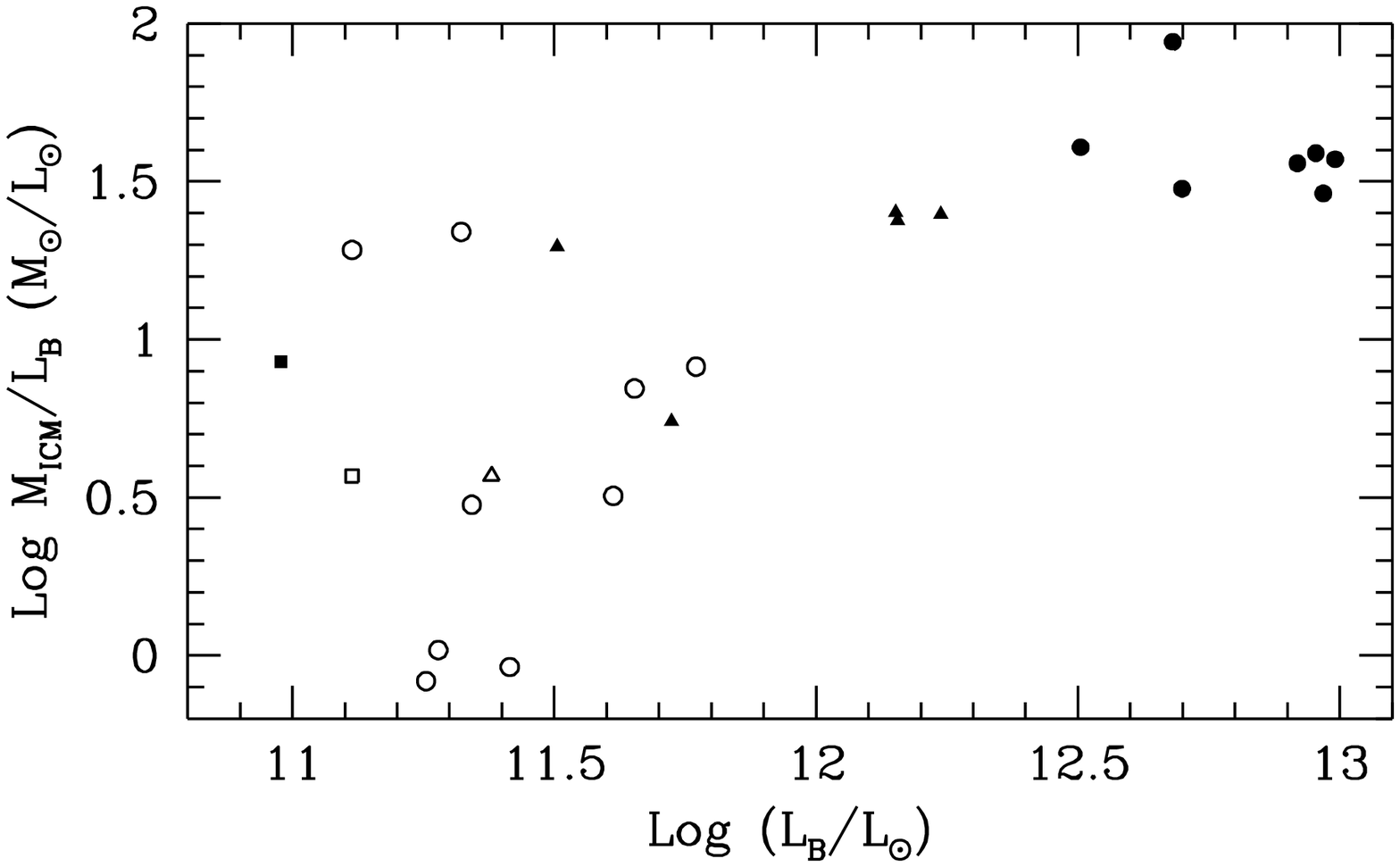]{The ICM mass to light ratio, i.e., the mass
of the ICM per
unit light of the cluster galaxies, as a function of the total optical
luminosity $\lb$ of the cluster galaxies. The same objects as in
Figure 1 are displayed.
\label{fig3}}

\figcaption[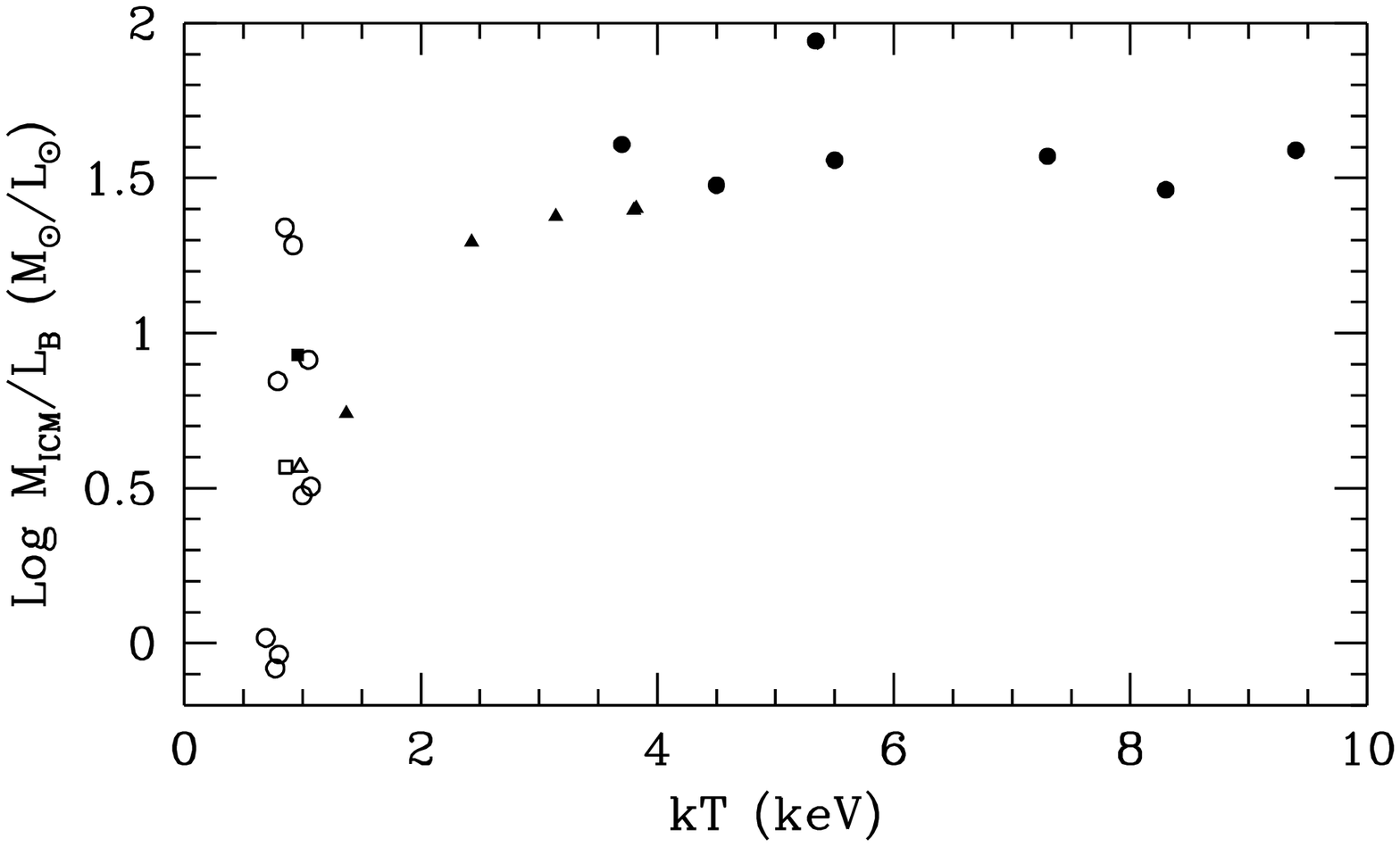]{The same as Figure 3 but as a function of
 the ICM temperature.
\label{fig4}}

\figcaption[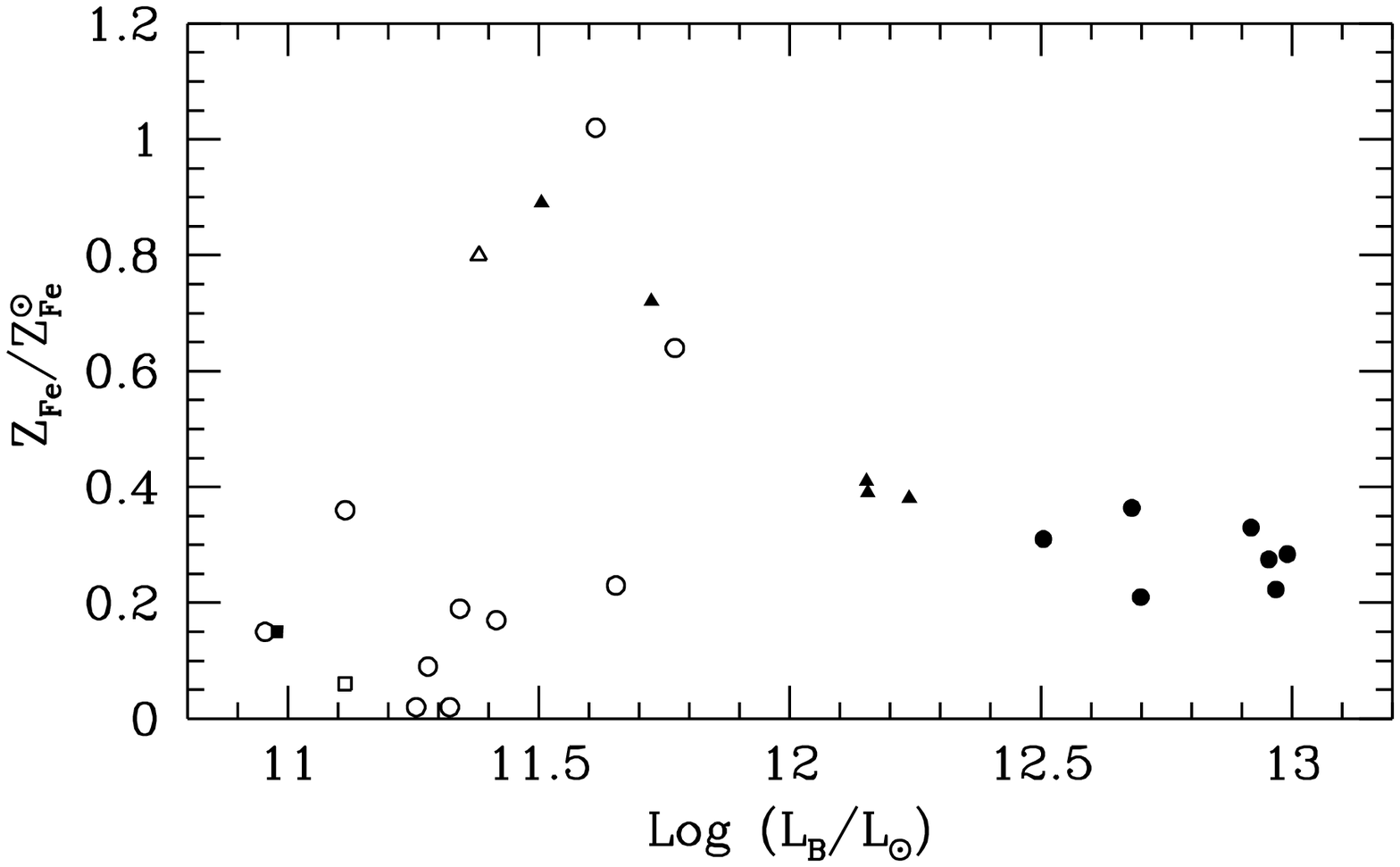]{The iron abundance in the ICM for the
objects in Figure 1 as a function of the total optical luminosity $\lb$ of the
cluster galaxies. 
\label{fig5}}

\figcaption[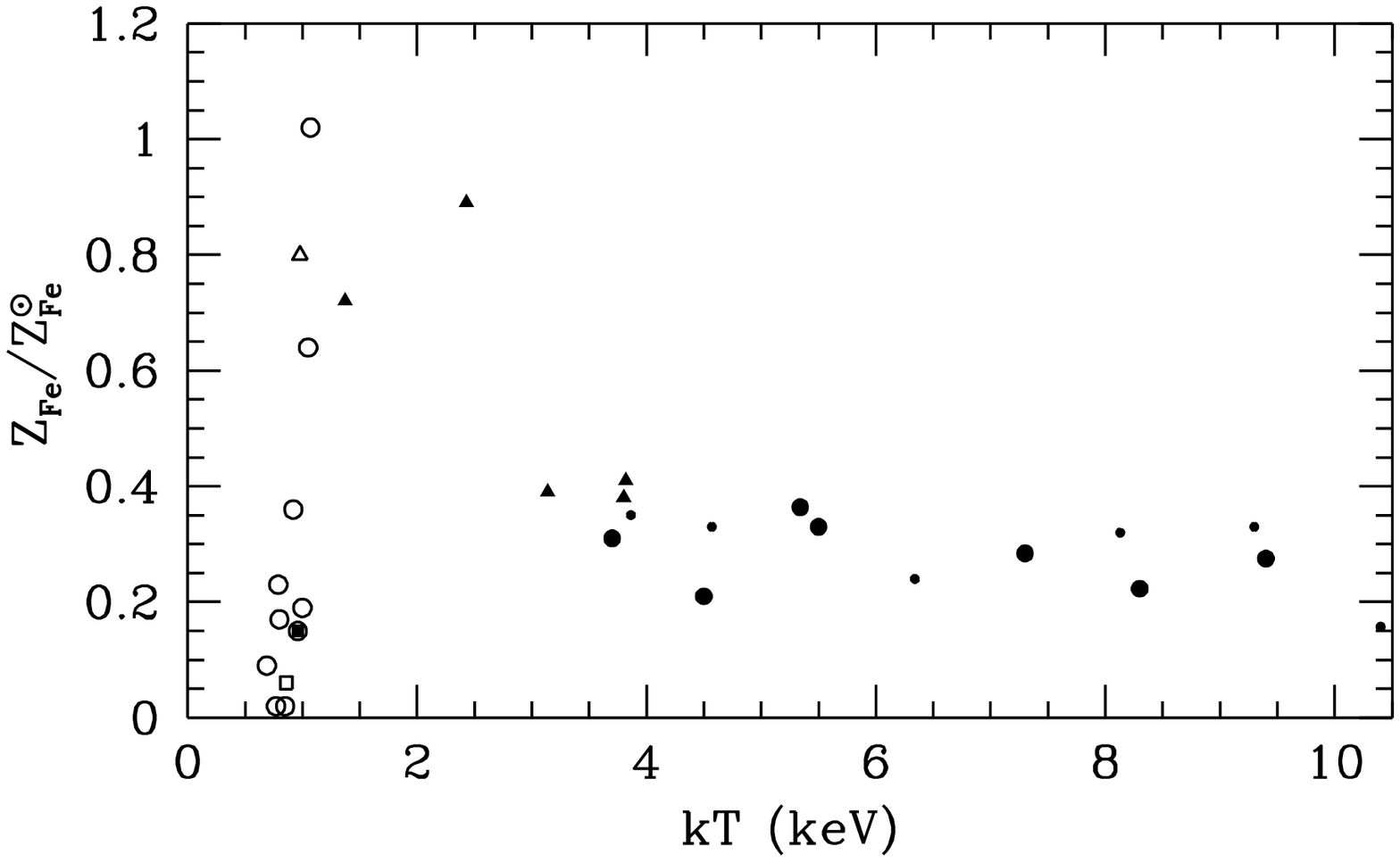]{The same as Figure 5 but as a function of
the ICM temperature.
Data for six clusters at moderately high redshift ($<z>\simeq 0.33$) 
are also included and represented by small filled circles 
(Allen et al. 1996; Donahue 1996; Matsuura et al. 1996; Schindler et
al. 1997). Note that for
$kT>\sim 3$ keV the iron abundance is derived from the iron-K complex,
while for lower temperatures the iron-L complex is used.
\label{fig6}}


\begin{thebibliography}{}
\bibitem []{} Anders, E., \& Grevesse, N. 1989, Geochimica et
     Cosmochimica Acta, 53, 197
\bibitem []{} Aragon-Salamanca, A., Ellis, R.S., Couch, W.J. \&
     Carter, D. 1993, MNRAS, 262, 764 
\bibitem []{} Arimoto, N., \& Yoshii, Y. 1987, A\&A, 173, 23
\bibitem []{} Arimoto, N., Matsushita, K., Ishimaru, Y., Ohashi, T., \&
    Renzini, A. 1997, ApJ, 477, 128 (AMIOR)
\bibitem []{} Arnaboldi, M., Freeman, K.C., Mendez, R.H., Capaccioli, M.,
          Ciardullo, R., Ford, H., Gerhard, O., Hui, X., Jacoby, G.H.,
          Kudritzki, R., \& Quinn, P.J. 1996, ApJ, 472, 145
\bibitem []{} Arnaud, M., Rothenflug, R., Boulade,O., Vigroux, R., \&
    Vangioni-Flam, E. 1992, A\&A, 254, 49 
\bibitem []{} Bender, R., Ziegler, B., \& Bruzual, G.A. 1996, ApJ, 436, L51
\bibitem []{} Bessell, M.S., Sutherland, R.S., \& Ruan, K. 1991, ApJ 383, L71
\bibitem []{} Bower, R.G., Lucey, J.R. \& Ellis, R.S. 1992, MNRAS, 254, 613
\bibitem []{} Canizares, C.R., Clark, G.W., Jernigan, J.G., \&
     Markert, T.H. 1982, ApJ, 262, 33
\bibitem []{} Carlberg, R.G., Morris, S.L., Yee, H.K.C., Ellingson,
     E. 1996, Astro-ph 9612169
\bibitem []{} Ciotti, L., D'Ercole, A., Pellegrini, S., \& Renzini,
    A. 1991, ApJ, 376, 380
\bibitem []{}  David, L.P., Arnaud, K.A., Forman, W., \& Jones, C. 1990, ApJ,
     356, 32
\bibitem []{} David, L.P., Jones, C., Forman, W., \& Daines, S. 1994, ApJ, 
     428, 544 
\bibitem []{} Davies, R.L., Sadler, E.M., Peletier, R.F., 1993, MNRAS,
     262, 650
\bibitem []{} Davis, D.S., \& White, R.E. III, 1996, ApJ, 470, L35
\bibitem []{} Dickinson, M. 1995, in Fresh Views of Elliptical Galaxies,
         ed. A. Buzzoni, A. Renzini, \& A. Serrano, ASP Conf. Ser. 86, 283
\bibitem []{} Elbaz, D., Arnaud, M., \& Vangioni-Flam, E. 1995, A\&A, 303, 345
\bibitem []{} Ellis, R.S., Smail, I., Dressler, A., Couch, W.J., Oemler,
     A. Jr., Butcher, H., \& Sharples, R.M. 1996, astro-ph
     9607154
\bibitem []{} Evrard, A.E. 1997, MNRAS, in press (astro-ph 9701148)
\bibitem []{} Ferguson, H.C., \& Davidsen, A.F. 1993, ApJ, 408, 92
\bibitem []{} Greggio, L. 1997, MNRAS, 285, 151
\bibitem []{} Greggio, L., \& Renzini, A. 1983, Mem. SAIt, 54, 311
\bibitem []{} Hui, X., Ford, H.C., Ciardullo, R., \& Jacobi, G.H. 1993, ApJS,
     88, 423
\bibitem []{} Hwang, U., Mushotzky, R.F., Loewenstein, M., Markert,
     T.H. Fukazawa, Y., \& Matsumoto, H. 1997, ApJ, 476, 560
\bibitem []{} Ishimaru, Y., \& Arimoto, N. 1997, PASJ, in press (astro-ph
     9702036)
\bibitem []{} Jablonka, P., \& Alloin, D. 1995, A\&A, 298, 361
\bibitem []{} Kaiser, N. 1991, ApJ, 383, 104
\bibitem []{} Kodama, T., \& Arimoto, N. 1997, A\&A, in press
      (astro-ph 9609160)
\bibitem []{} Larson, R.B. 1974, MNRAS, 169, 229
\bibitem []{} Lilly, S.J., Tresse, L., Hammer, F., Crampton, D., \& Le F\`evre,
      O. 1995, ApJ, 455, 108
\bibitem []{} Loewenstein, M., \& Mushotzky, R.F. 1996a, ApJ, 466, 695
\bibitem []{} --------------. 1996b, astro-ph 9608111
\bibitem []{} {Madau, P., Ferguson, H.C., Dickinson, M.E., Giavalisco, M.,
     Steidel, C.C.,\par \& Fruchter, A. 1996, MNRAS, 283, 1388
\bibitem []{} Matteucci, F. 1992, ApJ, 397, 32
\bibitem []{} Matteucci, F., \& Greggio, L. 1986, A\&A, 154, 279
\bibitem []{} Matteucci, F., \& Vettolani, G. 1988, A\&A, 202, 21
\bibitem []{} McWilliam, A., \& Rich, R.M. 1994, ApJS, 91, 749
\bibitem []{} Melnick, J., White, S.D.M., \& Hoessel, J. 1977, MNRAS, 180, 207
\bibitem []{} Mitchell, R., Ives, J., \& Culhane, L. 1975, MNRAS, 175, 29
\bibitem []{} Moore, B., Katz, N., Lake, G., Dressler, A., \& Oemler, 
       A. 1996, Nature, 379, 613
\bibitem []{} Mulchaey, J.S., Davis, D.S., Mushotsky, R.F., \&
       Burstein, D. 1993, ApJ, 404, L9
\bibitem []{} --------------. 1966, ApJ, 456, 80
\bibitem []{} Mushotzky, R.F. 1994, in Clusters of Galaxies, ed. F. Durret,
     A. Mazure, \& J. Tran Thanh Van (Gyf-sur-Yvette: Editions
     Fronti\`eres), p. 167
\bibitem []{} Mushotzky, R.F., Loewenstein, M., Awaki, H., Makishima,
      K., Matsushita, K., \& Matsumoto, H., 1994, ApJ, 436, L79
\bibitem []{} Mushotzky, R.F., Loewenstein, M., Arnaud, K., Tamura, T.,
      Fukazawa, Y., Matsushita, K., Kikuchi, K., \& Hatsukade, I., 1996,
      ApJ, 466, 686
\bibitem []{} Nath, B.B., \& Chiba, M. 1995, ApJ, 454, 604
\bibitem []{} Ohashi, T., Fukazawa, Y., Ikebe, Y., Ezawa, H., Tamura, T., \&
     Makishima, K. 1995, in  New Horizons of X-ray Astronomy, ed. F. Makino
     \& T. Ohashi (Tokyo: University Academic Press), p. 234
\bibitem []{} Ortolani, S., Renzini, A., Gilmozzi, R., Marconi, G., Barbuy, B.,
     Bica, E., \& Rich, R.M., 1995, Nature, 377, 701
\bibitem []{} Pahre, M.A., Djorgovski, S.G., \& de Carvalho, R.R. 1997, in
     Galaxy Scaling Relations: Origins, Evolution and Applications,
     ed. L. da Costa \& A. Renzini (Berlin: Springer), in press
\bibitem []{} Pei, Y.C., \& Fall, M. S.M. 1995, ApJ, 454, 69
\bibitem []{} Ponman, T.J., et al. 1994, Nature, 369, 462
\bibitem []{} Renzini, A. 1994, in Clusters of Galaxies, ed. F. Durret,
     A. Mazure, \& J. Tran Thanh Van (Gyf-sur-Yvette: Editions
     Fronti\`eres), p. 221
\bibitem []{} --------------. 1995, in Stellar Populations, ed. P.C. van der
     Kruit \& G. Gilmore (Dordrecht: Kluwer), p. 325
\bibitem []{} Renzini, A., \& Ciotti, L. 1993, ApJ, 416, L49}
\bibitem []{} Renzini, A., Ciotti, L., D'Ercole, A., \& Pellegrini, 
     S. 1993, ApJ, 419, 52 (RCDP)
\bibitem []{} Rosati, P., Della Ceca, R., Giacconi, R., \& Norman, C. 
     1997, Preprint
\bibitem []{} Ruiz-Lapuente,P., Burkert, A., \& Canal, R. 1996,  ApJ 447, L69
\bibitem []{} Serlemitsos, P., Smith, B., Boldt, E., Hold, S.S., \& Swank,
     J. 1976, ApJ, 211, L63
\bibitem []{} Trentham, N. 1994, Nature, 372, 157
\bibitem []{} Theuns, T., \& Warren, S.J. 1997, MNRAS, in press
\bibitem []{} Tsuru, T.  1993, PhD Thesis, University of Tokyo, ISAS RN 528
\bibitem []{} van Dokkum, P.G., \& Franx, M. 1996, MNRAS, 281, 985
\bibitem []{} Vigroux, L. 1977, A\&A, 56, 473
\bibitem []{} Wheeler, J.C., Sneden, C., \& Truran, J.W.Jr. 1989, 
      ARA\&A, 27, 279
\bibitem []{} White, S.D.M. 1997, in The Early Universe with the VLT,
     ed. J. Bergeron (Berlin: Spinger), p. 219
\bibitem []{} White, S.D.M., Navarro, J.F., Evrard, A.E., \& Frenk,
     C.S. 1993,   Nature, 366, 429
\bibitem []{} Woosley, S.E., \& Weaver, T.A. 1995, ApLS, 101, 181
\bibitem []{} Worthey, G., Faber, S.M., \& Gonz\'alez, J.J. 1992, ApJ, 398, 69
\bibitem []{} Worthey, G., Trager, S.C., \& Faber S.M. 1995, in Fresh Views of 
     Elliptical galaxies, ed. A. Buzzoni, A. Renzini, \& A. Serrano,
    ASP Conf. Ser. 86, 203
\bibitem []{} Zinnecker, H., McCaughrean, M.J., \& Wilking, B.A. 1993, in
    Protostars and Planets III, ed. E.H. Levy \& J.I. Lunine (Tucson:
    Univ. Arizona), p. 429
\end{thebibliography}
\end{document}